\documentstyle[preprint,prl,aps,epsf]{revtex}

\draft

\title{Single-Particle Diffusion-Coefficient on Surfaces with
  Ehrlich-Schwoebel-Barriers}  

\author{K. Mussawisade, T. Wichmann and K.W. Kehr}
\address{Institut f\"ur Festk\"orperforschung, Forschungszentrum
  J\"ulich GmbH,  D-52425 J\"ulich, Germany}

\begin{document}
\maketitle

\begin{abstract}

The diffusion coefficient of single particles in the presence of 
Ehrlich-Schwoebel barriers (ESB)is considered. An exact expression is given
for the diffusion coefficient on linear chains with random arrangements
of ESB. The results are extended to surfaces having ESB with uniform extension
in one or both directions. All results are verified by Monte Carlo simulations.

\end{abstract}
%\pacs{}

\section{Introduction}
Diffusion of particles on stepped surfaces is of great current
interest. Stepped surfaces occur in the form of more or less regular
terraces on vicinal surfaces, or in the process of island
formation. Ehrlich and Schwoebel \cite{ehrlich,swb1} pointed out the
existence and importance of what are now called Ehrlich-Schwoebel
barriers. A particle which  diffuses on a higher terrace 
of a stepped surface experiences a
higher than normal barrier at the  edge of the terrace while a
particle on the lower terrace experiences a lowered site energy when
approaching the step edge. In this paper the diffusion coefficient of
single particles on surfaces with
one- and two-dimensional arrangements of
Ehrlich-Schwoebel barriers will be calculated.  

In previous papers \cite{natori,meri} the diffusion coefficient of a
particle in a one-dimensional periodic arrangement of
Ehrlich-Schwoebel barriers has been derived. In Ref. \cite{meri} also
the collective diffusion coefficient of a finite concentration of
particles in periodic arrangements of Ehrlich-Schwoebel barriers has
been calculated in a mean-field approximation. We will present  results
for single-particle diffusion 
that are valid for disordered arrangements of such barriers. Our
derivations have been made possible by recent progress in the theory of
diffusion in
disordered lattices  \cite{MWK,kut,diet}. An exact result for the
diffusion coefficient of a particle on a linear chain with rather
general transition rates is now available and will be described in the
following  section. This result can be applied to the case of
Ehrlich-Schwoebel barriers and allows for a simple and exact treatment
of both ordered and disordered arrangements of the barriers.
An important point is the
extension to two-dimensional surfaces which will be done in
Sec.III. The final section contains concluding remarks.

\section{Diffusion coefficient on linear
chains with Ehrlich-Schwoebel-Barriers}

The asymptotic behavior of the mean-square displacement of independent
particles is characterized by a diffusion coefficient which is given
for linear chains with rather arbitrary transition rates by
(\cite{MWK,kut,diet}) 
\begin{equation} D = \left\{ \frac{1}{\rho_i
      \Gamma_{ij}} \right\}^{-1} \label{dex1} \:. 
\end{equation}
$\rho_i$ is the thermal occupation factor of site $i$, $\Gamma_{ij}$ the
transition rate from site $i$ to site $j$, and the lattice constant has been
set unity. The curly brackets designate the disorder average which has
to be taken over the distributions of the site and barrier energies. 
The thermal occupation factors
are defined by
\begin{equation}
\rho_i = \frac{\exp (-\beta E_i)}{\{\exp (-\beta E_i)\}}  \label{occ}
\end{equation}
where $E_i$ is the energy of site $i$. The transition rates are given by
an Arrhenius expression,
\begin{equation}
 \Gamma_{ij} = \Gamma_0 \exp (-\beta (E_{ij} - E_i)) \label{jr}
\end{equation}
where $E_{ij}$ is the energy barrier between site $i$ and site $j$,
and $\beta$ is the inverse temperature (i.e. $\beta = 1/k_B T$). The
site energies $E_i$ , counted from a common origin, shall always be
nonpositive while the barrier energies $E_{ij}$ are assumed to be
nonnegative. In this way all transition rates fulfill $\Gamma_{ij} \le
\Gamma_0$.  We emphasize that the expression (\ref{dex1}) for the 
diffusion coefficient  is only valid for systems which have a unique
equilibrium state in the limit of number of sites $N \rightarrow
\infty$; otherwise the thermal occupation factors $\rho_i$ are not
defined. 

The Ehrlich-Schwoebel barriers (ESB), which have been postulated from
experimental observations, are of the type where Eq.(\ref{dex1}) is
applicable. A pictorial  representation of ESB on a linear chain is
given in Fig.I in the form of a $1$-dimensional potential
landscape. The potential has to transcribed  to transition rates by
application of the Arrhenius expression (\ref{jr}). According to
Eq.(\ref{dex1}) an average over the inverse of weighted transition
rates has to be performed.  Multiplying the occupation factors
(\ref{occ}) with the rates (\ref{jr}) one  observes a cancellation of
the factors $\exp(-\beta E_i)$. An effective  factorization of the
expression results and it reads  
\begin{equation}
  D = \left\{\frac{1}{\exp(-\beta E_i)}\right\}_{E_i}^{-1} 
  \left\{\frac{1}{\exp(-\beta E_{ij})}\right\}_{E_{ij}}^{-1},
  \label{fact}
\end{equation} 

Now the diffusion coefficient of a particle on a chain with randomly
distributed ESB of identical heights and depths
(see Fig 1) is easily calculated, 
\begin{equation}
 D =\frac{1}{\left(1 - c_s +  c_s \exp(\beta E_s)\right)
     \left(1 - c_s + c_s \exp(-\beta E_t)\right)} \label{dswb}
\end{equation}
where $c_s$ is the concentration of the 
ESB, $E_s$ the height of the ESB and $E_t$ the energy of the site
at a surface step. 
The result (\ref{dswb}) is equivalent to the results derived in
\cite{natori} and \cite{meri}. The authors of \cite{natori} and
\cite{meri}  consider regularly stepped surfaces i.e. the length
$L$ between two steps is always constant. In our result the diffusion
coefficient only depends on the concentration $c_s$ of 
the ESB and $L$ need not to be constant, i.e. $c_s = 1/\{ L \}$
where $\{ L\}$ is the average distance of the barriers. 

In (FIG 2.) one recognizes good agreement of Monte Carlo results
for particle diffusion on such chains with the diffusion coefficient 
as given in Eq. (\ref{dswb}).

A comment on the absence of directional effects in the diffusion 
coefficient on linear chains with ESB is in order. Since the 
diffusion coefficient of the linear chain with ESB has
the factorized form eq.(\ref{fact}), the arrangements of the 
high barriers and the deep sites does not matter for the 
magnitude of it. This seems to contradict the idea that the 
ESB prevent particles from flowing from higher terraces to lower
ones. We point out that we derive the diffusion coefficient of
single, independent particles, which does not exhibit any 
directional effects (see also \cite{natori,KMWD}). In the linear-response
regime the mean flux of single particles up and down the steps 
is the same for a given small bias. The ESB become effective for
the formation of islands, if there are several particles present on
the highest terrace. 
If one particle is kept next to a high barrier at the terrace edge,
a second particle may attach to this particle, and so on,
leading eventually to the formation of a higher terrace.   

In the paper Ref. \cite{meri} collective diffusion on stepped 
surfaces with ESB was considered for lattice gases with exclusion
of double occupancy. No further interactions of the particles were
considered. The formula  Eq.(4) in \cite{meri} for $D_{xx}$ agrees in 
the limit $c \rightarrow 0$ with the previous result \cite{natori}
and our result. The limiting behavior for $c \rightarrow 1$ can be  
directly studied by our methods. The problem is then equivalent to a 
single-vacancy diffusion problem where the transition rates {\sl into}
the sites are given \cite{KPW}. Application of Eq.(\ref{dex1}) leads to
\begin{equation}
  \label{cto1}
D(c \rightarrow 1) = [(1-c_s + c_s\exp (-E_t))((1-c_s + c_s\exp(\beta
(E_t+E_s)) +c_s\exp(\beta E_t))]^{-1}
\end{equation}
The expression given in \cite{meri} gives an interpolation between
$c=0$ and $c=1$ and approaches for $c \rightarrow 1$ the result 
Eq.(\ref{cto1}). We emphasize that in the site-exclusion lattice-gas
model which was used in \cite{meri} additional interactions of the 
particles are neglected. Interaction effects of the particles are 
expected to be important for real surfaces, hence the 
applicability of the formulae to real systems needs careful examination.

\section{Diffusion coefficient on two-dimensional surfaces}
 
The theory of diffusion in the presence of ESB can be extended to 
two-dimensional surfaces under the assumption that the random walk of
the particle(s) can be considered as independent in the two surface 
directions. There are two cases where this condition is fulfilled: 

a) {\sl Terraces with infinite uniform extension in one direction}\\
This case is of practical interest because vicinal surfaces may have
steps which extend uniformly in the direction perpendicular to the 
steps. See Fig.3 for a pictorial representation. Clearly the random 
walk of a particle in the  x- and y-directions can be considered as
independent. 
While the particle experiences ESB barriers when it performs random
walk in the x-direction, it can make jumps in the y-direction with
a uniform, site-independent transition rate $\Gamma$. The summary 
diffusion coefficient is then given by  
\begin{equation}
  \label{twod}
  D = \frac{1}{2}( D_x + D_y) \:.
\end{equation}
The quantity $D_x$ is given by (\ref{dex1}) while $D_y$ is simply
$D_y = \Gamma$ when the lattice constant is set unity. Fig.4 shows the 
result of simulations for such a 2-dimensional model, together with the
predictions of Eq.(\ref{twod}). One notices complete agreement of theory 
and simulations. 

In Ref.\cite{natori} also the case was considered where the transition
rates between the sites with lowered energies at the
edge of the terrace, in $y$-direction, have a smaller value, $\Gamma_2$. The result for the 
perpendicular diffusion coefficient $D_y$ is then a superposition of 
the contributions of the different rows in $y$-direction. The result
of \cite{natori} for $D_y$ is easily extended to the case of varying
terrace distances $L$.

b) {\sl Independent uniform terraces in x and y directions}.\\
If there are independent uniform terraces in x and y direction, for example as
indicated in Fig.5, the diffusion coefficient may again be calculated 
by applying the principle of independent random walks in the two directions.
It is assumed that ESB exist at the terrace steps in each direction. The
diffusion coefficient $D$ is again obtained from the superposition
Eq.(\ref{twod}), but now $D_y$ is also given by the result
Eq.(\ref{dswb}). Figure 6 demonstrates the verification of the
analytical result by numerical  simulations. 

\section{Concluding remarks}

We have derived the diffusion coefficient of single particles on linear 
chains with random arrangements of ESB. We could extend the results 
to two-dimensional surfaces with terraces and associated ESB with 
infinite uniform extension in one or two directions. All results were 
validated by Monte Carlo simulations. Recently also the necessity of
introducing more complicated barriers at the step edges, than the
conventional ESB, has been pointed out \cite{KE}. The derivation of the
diffusion coefficient for such barrier structures from our
Eq. (\ref{dex1}) is straight forward. One open problem is the treatment
of diffusion of single particles on terraces which do not extend 
straightly in the perpendicular direction, i.e., terraces with  
kinks or more complicated defects. On stepped surfaces the diffusion 
can only be separated in 
contributions of each direction if the probability of a jump parallel
to the steps does not depend on the position of the particle regarding
to the steps. Thus the single-particle diffusion 
coefficient is not yet known for more complicated forms of terraces. 
Another unsolved problem is the diffusion coefficient of many 
interacting particles
which obey site exclusion (i.e., interacting lattice gases), on
stepped surfaces  
with ESB. The solution of this problem would be important for an 
analytic understanding of ,e.g., island growth.

We thank  J. Krug, T. Michely and M. Rost for discussions. 

%\begin{figure}[htp]
%  \epsffile{mobopt.eps}
%  \caption[]{} 
%\end{figure}

\newpage

\begin{center}
{\bf Figure Captions}
\end{center}

\noindent FIG.\ 1. Ehrlich-Schwoebel barriers

\vskip 0.5cm

\noindent FIG.\ 2. Diffusion coefficient of a particle vs. concentration of the
  barriers  for different Temperatures in the 1-dimensional case. Lines: exact
  calculation. Points Monte Carlo Simulations. 
\vskip 0.5cm

\noindent FIG.\ 3 Ehrlich-Schwoebel barriers extended to two dimensions.

\vskip 0.5cm

\noindent FIG.\ 4. Diffusion coefficient of a particle vs. concentration of the
  barriers for different Temperatures in the 2-dimensional case where
  the terraces are uniformly extended in the $y$-direction. Lines:
  exact calculation. Points Monte Carlo Simulations.

\vskip 0.5cm

\noindent FIG.\ 5 Ehrlich-Schwoebel barriers in both $x$ and $y$-directions.

\vskip 0.5cm

\noindent FIG.\ 6 Diffusion coefficient of a particle vs. concentration of the
  barriers in the 2-dimensional case where in both $x$ and
  $y$-directions barriers occur (cf. Fig. 5). Lines: exact calculation
  for different  Temperatures. Points: Monte Carlo Simulations.

\begin{figure}[htp]
  \epsffile{swblandscape.eps}
  \caption[]{} 
\end{figure}

\begin{figure}[htp]
  \epsffile{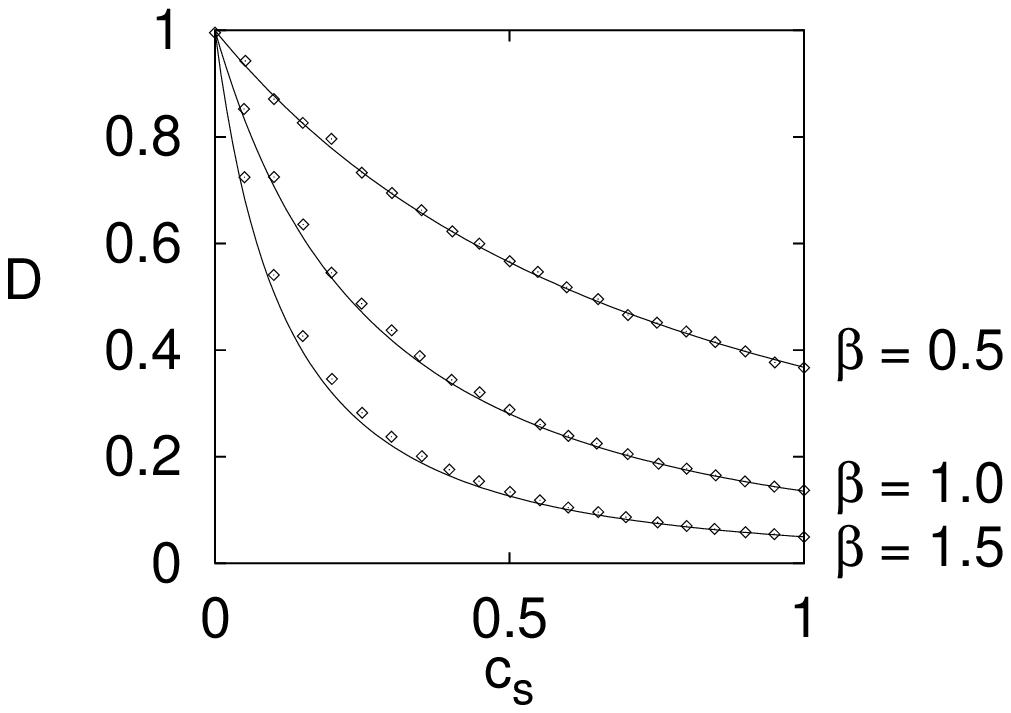}
  \caption[]{} 
\end{figure}

\newpage

\begin{figure}[htp]
  \epsffile{swbd2fig2.eps}
  \caption[]{} 
\end{figure}

\begin{figure}[htp]
  \epsffile{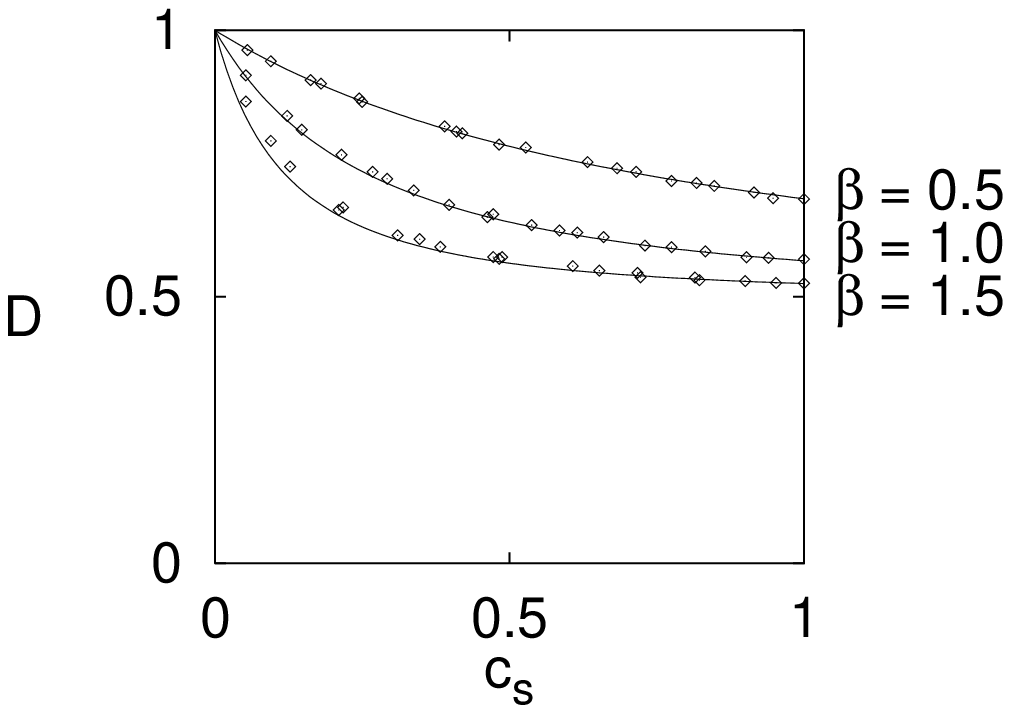}
  \caption[]{} 
\end{figure}

\newpage

\begin{figure}[htp]
  \epsffile{swbxy.eps}
  \caption[]{} 
\end{figure}

\begin{figure}[htp]
  \epsffile{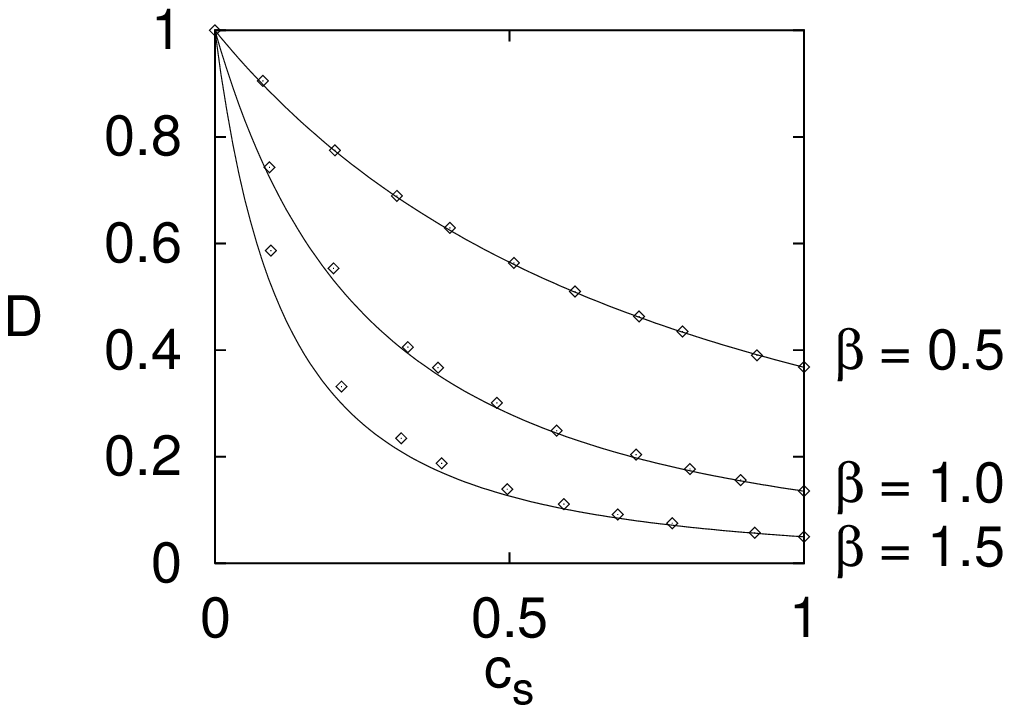}
  \caption[]{} 
\end{figure}

\end{document}